# Origin of Robust Rectification in Geometric Diodes

Mengmeng Bai, Yanqing Zhao, Shuting Xu, Yao Guo*

*School of Physics, Beijing Institute of Technology, Beijing 100081, China*

**Abstract:** Geometric diodes, which take advantage of geometric asymmetry to achieve current flow preference, are promising for THz current rectification. Previous studies relate geometric diodes' rectification to quantum coherent or ballistic transport, which is fragile and critical of the high-quality transport system. Here we propose a different physical picture and demonstrate a robust current rectification originating from the asymmetric bias induced barrier lowering, which generally applies to common semiconductors in normal environments. Key factors to the diode's performance are carefully analyzed, and an intrinsic rectification ability at up to 1.1 THz is demonstrated.

Diode, the two-terminal electronic component that rectifies current in one direction, is among the simplest but most useful semiconductor devices [1-4]. *p-n* junction diodes and Schottky junction diodes are the most common solid-state rectifier diodes, which are widely used in power and signal processing circuits for AC-DC conversion, envelope detection, amplitude clipping, voltage regulation, etc. [5-8]. *p-n* junction diodes have excellent rectification characteristics and are high-voltage tolerant. However, as the minority carrier device, the *p-n* junction diodes suffer from the minority charge storage, which hinders their high-frequency rectification performance [9-11]. Schottky diodes are of particular interest as detectors, mixers, and frequency multipliers for high-frequency rectification because of their relatively high switching speeds. However, metal-semiconductor junctions are volatile and incapable of working at a high-voltage range [12-14].

Besides the *p-n* junction diodes and Schottky junction diodes, geometric diodes that take advantage of their geometric asymmetry were raised in the late 1990s [15-17]. The geometric diodes' rectification have been related to quantum coherent or ballistic transport, which are fragile and critical of the transport system. Therefore, geometric diodes had been only demonstrated on high-mobility III-V semiconductor materials and graphene [18-25]. In a recent paper, Custer et al. reported a silicon nanowire based geometry diode, which enabled robust current rectification at a room temperature [25]. The current rectification was attributed to a ratchet-like quasi-ballistic transport in diodes. Here we keep the reservation about the ratchet-like quasi-ballistic model, which might be over-idealized for doped vapor-liquid-solid grown silicon nanowires at room temperature. The experimentally observed rectification ratio was far beyond the theoretical expectation, and it is unreproducible in the Monte-Carlo carrier transport simulation [see detailed arguments in Supplemental Material I and Figs. 1(a)].

In this work, we investigate the origin of the robust rectification in the geometric diodes. We demonstrate a robust current flow preference originated from an asymmetric bias-induced-barrier-lowering (BIBL), where the ballistic transport or quantum coherent is not a necessity. The current rectification is well reproducible using a basic electrostatic semiconductor physics model and shows good accordance with the experimental observations. Vital factors to the geometric diode, such as the diode geometry, surface charging, passivation, and temperature, are carefully analyzed. The geometric diode demonstrates an intrinsic rectification ability at up to 1.1 THz. This work provides a fundamental comprehension of geometric diodes, revealing their essential physical mechanisms and extending their applicable range from specific high-mobility candidates to all general semiconductor matters.

The model is built in a technology computer-aided design environment. As is shown in Fig. 1(a), the geometric diode features an asymmetrical bottleneck. Here we set the geometric parameters and doping profile according to Ref. [25]. Our initial simulation did not show significant current rectification, as is shown in the inset diagram of Fig. 1(b). The absence of current rectification is in accordance with the finite-element simulation in Ref. [25]. However, we note that such models neglected the surface states, which are well-known to exist at the surface of silicon. We added the surface states ($2.5 \times 10^{12}$ cm$^{-2}$, acceptor-like) to the model, and the *I-V* curve show an evident current rectification ratio of 67, as shown in Fig. 1(b). Therefore, a basic electrostatic semiconductor device model can well reproduce the rectification behavior in the geometric diode. The simulated band diagrams are shown in the three lower extensive diagrams of Fig. 1b and Figs. 2(a-f). Apparently, a potential barrier exists along the axial direction of the diode. The barrier height varies with the applied voltage, as summarized in Fig. 1(c). The unbiased barrier height is 471 meV, which can be lowered by the applied bias voltage asymmetrically: the forward bias lowers the barrier more significantly, with the barrier height decreased to 48 meV at 1 V, while the reverse bias lowers the barrier less significantly, with the barrier height decreased to 193 meV at -1 V. The potential barrier, which can be asymmetrically lowered by the bias, is obviously determinative to the carrier transport in the geometric diode.

We now expound on the mechanism of the geometric diode's robust rectification behavior. As shown in Fig. 2(a), the terminated atomic lattices at the surface generate surface states [26-28]. These surface states can give rise to the Fermi level pinning, making the narrow bottleneck fully depleted (see the Figs. 3 in supplementary) [29,30]. The full depletion region forms a potential barrier, which blocks the majority carrier (electron for the *n* doped channel in this case) transporting from one side to the other. As shown in Fig. 2(c-e), the potential barrier height can be lowered by the bias, which is akin to the well-known drain-induced-barrier-lowering (DIBL) in short channel field-effect transistors [31-33]. Due to the asymmetric geometry and the intense fringe field-effect at the kink in Fig. 2(a), the BIBL is asymmetric. For forwarding bias (see upper in Fig. 2(b)), the fringe electrical field orients to the channel, depletes the carriers, and lowers the barrier height more significantly. In contrast, for reverse bias (see lower in Fig. 2(b)), the fringe electrical field orients from the channel, accumulates the carriers, and prevents the barrier height from lowering. The asymmetric BIBL results in the current flow preference of the diode, according to the Boltzmann distribution. Ideally, we have

$$\frac{I_+}{I_-} \propto exp\left(\frac{\Phi_- - \Phi_+}{kT}\right) \qquad (1)$$

where $I_+$ and $I_-$ are the forward and reverse current, respectively, $k$ is the Boltzmann constant, $T$ is temperature, $\Phi_+$ and $\Phi_-$ are the potential barrier heights at the forward and reverse bias. Therefore, the rectification can be well explained based on fundamental semiconductor physical principles. Note that our analysis does not exclude the ballistic transport or quantum coherent induced rectification, though, as has been demonstrated in graphene and III-V component geometric diodes. However, such rectification is frail and critical of a high-quality transport system [18-24]. The BIBL induced rectification is robust and applies to a wide range of semiconductors without critical environmental requirements. We also note that for BIBL raised current rectification, the minor carrier (hole, in this case, see Figs. 4 and Figs. 6) density is not significantly increased. The absence of minority storage is vital for high-frequency rectification of the diode, as will be demonstrated below.

Based on the above comprehension, we now investigate key factors determining the geometric diode's rectification behavior.

**(1)** Surface states. In the above discussion, surface states are necessary to deplete the doped channel and generate a potential barrier at the bottleneck. As is shown in Fig. 3(a), our simulation results do show that for the doped nanowire geometric device, with increased surface state density, the current rectification increases, reaching 7000 with a surface state density of $3.5\times10^{12}$ cm$^{-2}$ at room temperature ($\varphi$=46°, $\theta$=5.29°, $D$=100nm, $d$=10nm). However, it should be noted that if the bottleneck of the nanowire is undoped, a potential barrier intrinsically exists, and the surface states are not necessary. A surface state free geometric diode can provide more stable electrical characteristics. The rectification ratio of a surface state free geometric diode reaches $7\times10^5$, as is shown in Fig. 3(b) and Supplemental Material V.

**(2)** Geometry: as shown in Fig. 3(c) and Figs. 7, we simulated geometric diodes with various φ of 25°, 45°, 64°, 64.5°, and 65°. An increased φ provides a more intense field-effect at the kink, resulting in the asymmetric BIBL. We observe a sudden increase of rectifying ratio at φ > 64°, which is in good accordance with the experimental observation Ref. [25].

**(3)** Passivation: The fringe field-effect and asymmetric BIBL can be further strengthened by dielectric coating. Fig. 3(d) and 3(e) show the *I-V* curve with and without the surface passivation layer of 10 nm SiO$_2$ and 10 nm Al$_2$O$_3$, respectively. The surface state density is set as low as $4\times10^{11}$ cm$^{-2}$. For the bare Si nanowire geometric diode, such low surface density cannot cause distinct current rectification (black line). For the passivated, we observe the apparent current rectification ratio of 9 (SiO$_2$) and 28 (Al$_2$O$_3$), respectively. Therefore, surface dielectric passivation can significantly enhance asymmetric BIBL and the current rectification of the geometric diode.

**(4)** Temperature: according to Eqs. (1), the rectification ratio monotonically decreases with temperature. The simulated temperature-dependent *I-V* curves of the geometric diode, as shown in Fig. 3(f). The rectification ratio dramatically increases with decreased temperature. Note that both the reverse and forward current were diminished as the temperature decreases, which is in accordance with the experimental observation because fewer carriers can hop over the barrier at a lower temperature.

Diodes working at the THz range are essential for the high throughput signal processing and communication technology. Due to the absence of minority carrier storage, geometric diodes are promising to achieve higher frequency current rectification over *p-n* junction diodes and even Schottky junction diodes. Here we present the high-frequency response of the geometric diode. The DC *I-V* curve of the geometric diode is shown in Figs .8(a), with a DC rectification ratio (± 1V) of 10. Such rectification maintains at an AC stimulation with frequency up to around 100 GHz, as shown in Fig. 4(a) and Figs. 8(b-j). The rectification ratio drops significantly at a higher frequency. Still, a rectification ratio of 2 is obtained at 1.1 THz, indicating that the diode can provide effective current rectificaton at the THz range, as shown in Fig. 4(c). The geometric diode further loses its rectification ability at higher frequency, and the rectification ratio decreases down to 1.1 at 2 THz, as shown in Fig. 4(d). By comparison, both *p-n* junction and Schottky junction silicon nanowire diodes lose their rectification at a lower frequency, as shown in Figs. 9(a) and (b), respectively. Therefore, the geometric diode has significant advantages over the *p-n* junction and Schottky junction diode at the THz range, showing great potential in ultra-fast signal processing circuits, 6G communications, and THz radar and electromagnetic wave energy harvesting.

**Conclusion**

In summary, this work proposes a robust rectification mechanism in the geometric diode. The robust rectification intrinsically originates from the asymmetric bias induced barrier lowering on the basis of fundamental semiconductor device physical principles. Therefore, such robust rectification applies not critically to the high mobility candidates but generally to all semiconductors. Key factors to the geometric diode such as surface states, geometry, surface passivation, and temperature are carefully analyzed, and the intrinsic rectification ability at THz range is demonstrated. This work provides a fundamental comprehension of the geometric diode, which shows excellent potential in THz circuits over the traditional *p-n* junction and Schottky junction diodes.

Acknowledgements: We thank Prof. James F. Cahoon, the author of Ref. [25], for the response of our queries. We also thank Tao Tang from Advanced Manufacturing EDA Co. Ltd for the help and useful discussions. This work is supported by the National Natural Science Foundation of China grant 11804024.

*Email: yaoguo@bit.edu.cn

# Figures

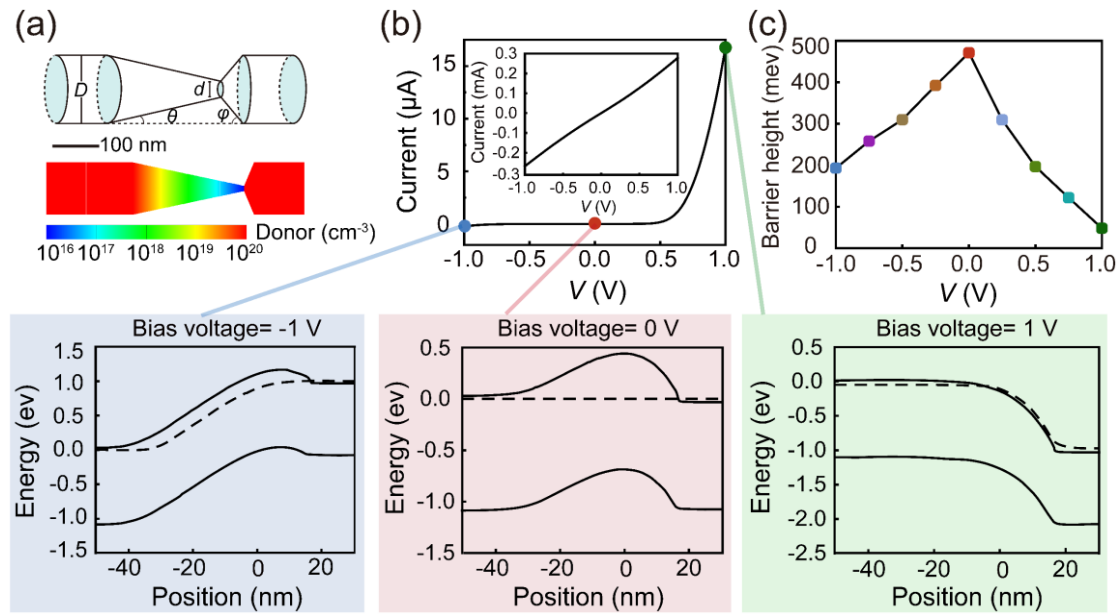

**FIG. 1. Electrostatic model of a geometric diode. (a)** Upper: geometric parameter definition, outer wire diameter (*D*), constriction diameter (*d*), angle (*θ*), and constriction angle (*φ*). Lower: device geometry and donor distribution. $D$=100 nm, $d$=10 nm, $θ$=5.29°, $φ$=46°. Scale bar: 100 nm. **(b)** Inset: initially simulated *I-V* curve of silicon nanowire diode without surface states. Main: simulated *I-V* curve with a surface density of $2.5×10^{12}$ cm$^{-2}$. Extended: The band diagram with the applied bias voltage of -1 V, 0 V, 1 V. **(c)** The barrier heights along the geometric diode axis with the various voltage.

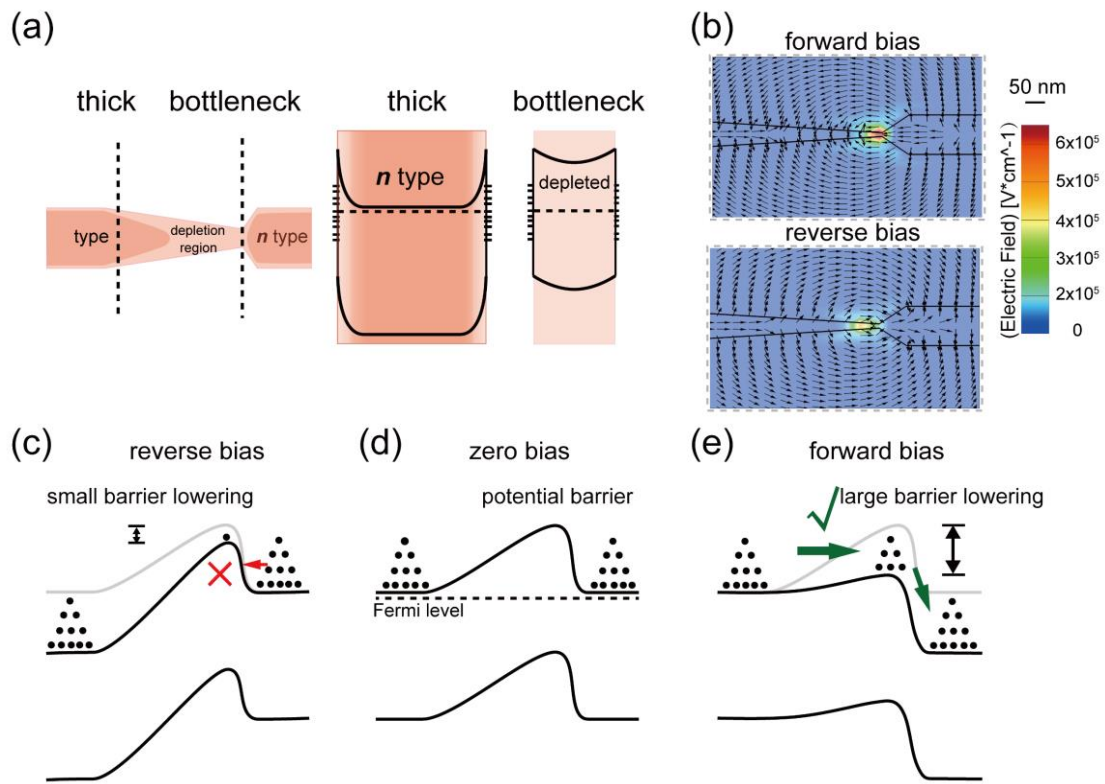

**FIG. 2. Schematic diagram for the origin of the rectification. (a)** Surface states, Fermi level pinning, depletion, and fringe field-effect. **(b)** The fringe electrical field distribution at bottleneck of silicon nanowire geometric diode at forward bias and reverse bias. **(c-e)** The band profiles of the diode at reverse **(c)**, zero **(d)**, and forward bias **(e)**.

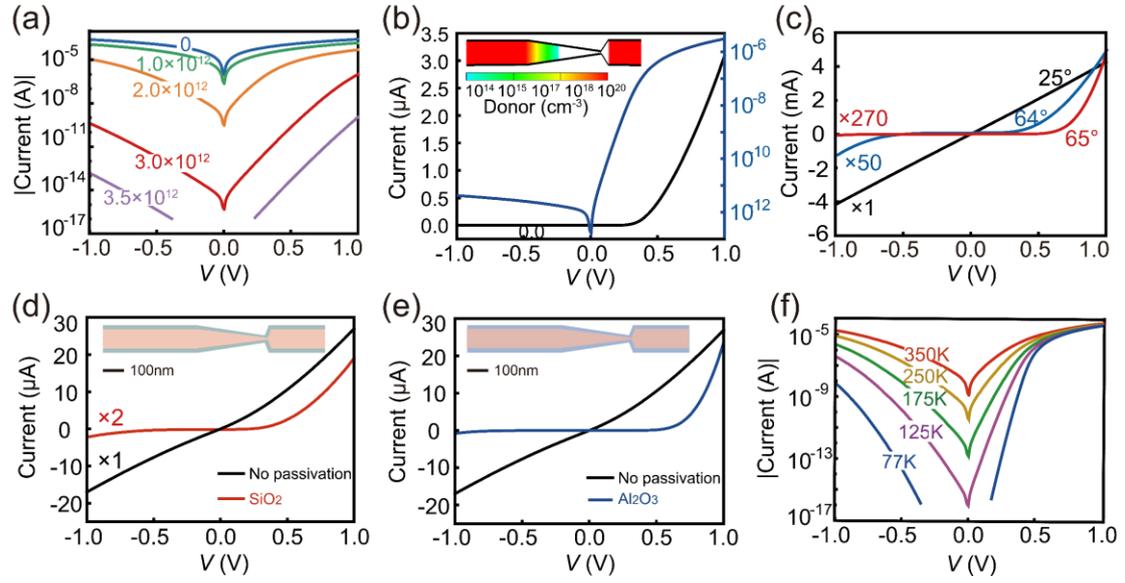

**FIG. 3. Key factors in geometric diodes.** (a) Log scale *I-V* curves of five geometric diodes with the surface states of 0, $1\times10^{12}$, $2\times10^{12}$, $3\times10^{12}$, $3.5\times10^{12}$ cm$^{-2}$ at room temperature. The current decreases, and the rectification ratio increase with higher surface density. (b) The *I-V* curve of geometric diodes with an undoped constrictive bottleneck. (c) The *I-V* curves simulated from three separate geometric diodes with $\varphi$ of 25°, 64°, 65°, respectively. The surface charge density was set $1.3\times10^{12}$ cm$^{-2}$. (d and e) The *I-V* curves from the geometric diode wrapped with 10 nm SiO$_2$ (d) and 10 nm Al$_2$O$_3$ (e), respectively. The charged surface state density was set at $4\times10^{11}$ cm$^{-2}$. (f) Log scale *I-V* curves of a geometric diode device at temperatures of 77 K (blue), 125 K (purple), 175 K (green), 250 K (yellow), and 350 K (red).

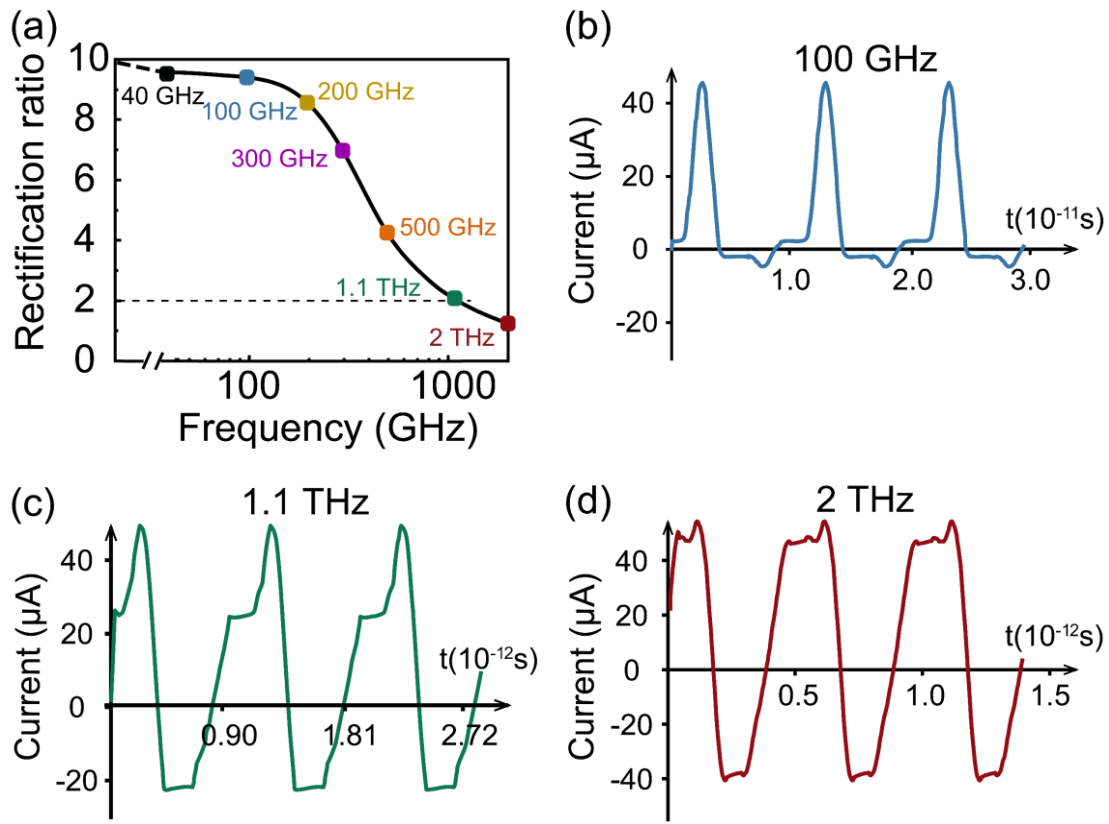

**FIG. 4. High-frequency response of the diodes. (a)** The rectification ratio of the geometric diode with frequency varying from 40 GHz to 2 THz. **(b-c)** The current *v.s.* time curve of the geometry diode at 100 GHz **(b)**, 1.1 THz **(c)**, 2 THz **(d)**, respectively.

Supplemental Material for:

# Origin of Robust Rectification in Geometric Diodes

Mengmeng Bai, Yanqing Zhao, Shuting Xu, Yao Guo*

*School of Physics, Beijing Institute of Technology, Beijing 100081, China*

# I. Arguments on the quasi-ballistic transport model

In recent work, Custer et al. demonstrated a vapor-liquid-solid (VLS) grown Si nanowire geometric diode with the largest rectification ratio of 1600 at room temperature[1]. The rectification was attributed to the quasi-ballistic transport of the carrier in Si. We argue, however, that the theory in [1] mismatches the experimental observations:

**(1) Over-idealized model and exceptionally underestimated rectification.**

In [1], it was proposed that current rectification results from the quasi-ballistic transport of the carrier. In fact, an 100% ballistic transport model was used in [1] assuming 100% specular scattering at the surface and 0 carrier scattering in bulk, which is obviously over-idealized for the doped VLS grown Si nanowire. Still, the largest rectification ratio derived from the theoretical model fit with the fabricated device geometry ($\varphi=65°$) is less than 1.5 (see Figure. 3C of [1]). In the experimental, however, the room temperature rectification ratio reaches ~10 and ~1600 (see Figure. 3A and 3B of [1]), the rectification ratio at 77 K reaches ~500 (see Figure. S6 of [1]), respectively. The vast mismatch between the theory and experimental results is obviously non-negligible.

The Monte-Carlo method is generally used to reproduce the quasi-ballistic/ballistic transport behavior in nanoscale electronic devices. As shown in Figs. 1, we simulate the transport of silicon geometric diodes in a Monte-Carlo model. Despite our multiple trials on various conditions, the simulated results cannot exhibit the rectification characteristics in [1] without introducing the BIBL and field-effect. Therefore, we do not find substantial evidence supporting that the quasi-ballistic carrier transport can generate the giant DC asymmetry observed in the experiment.

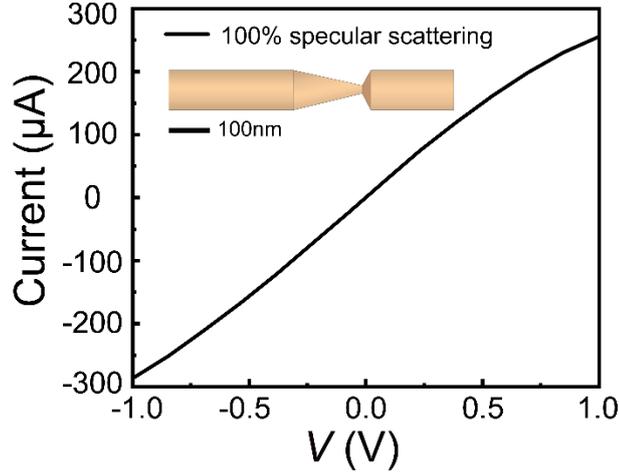

**FIG. S1.** *I-V* curve of the geometric diode from the Monte Carlo simulation. The surface scattering was set 100% specular. Scale bar: 100 nm.

Other arguments include:

**(2) Exceptionally small current in the diode**

In a quasi-ballistic (or ballistic) transport model, the conductance of the channel is $G = 2M \cdot T_C \cdot e^2/h$, where $M$ is the number of modes in the transmission channel, $e$ is the elementary charge, $h$ is Planck's constant, $T_C$ is the transmission probability, and $2e^2/h$ is the single-mode quantum conductance, which is 78 µA/V. The bottleneck of the geometric diode (~10 nm) should contain multiple channels, and the conductivity of the geometric diode can be larger than mA/V if most carriers transport ballistically through the bottleneck. However, in the experiment, the forward current is no more than several µA. For some geometric diodes, the forward current is as small as several $10^{-10}$ A (see Figure. S7 of [1]). The exceptionally small forward current indicates that the carriers are mostly blocked rather than going through ballistically.

### (3) Dramatically decreased current at low temperature

[1] claim that at low temperatures, the mean free path increases, and therefore the DC asymmetry increases. The increased mean free path should also make the diode more conductive, and increases the forward current. In the experiment, however, the forward and backward current decreased nearly 100 times and 3 times respectively as the temperature decrease from 350 K to 77 K. The dramatically decreased current with lower temperature is usually seen as evidence for the existence of the potential barriers [2].

### (4) Overestimated mobility in VLS silicon nanowire

[1] estimated the mobility and mean free path of their silicon nanowire according to the data of the bulk silicon. In a VLS grown silicon nanowire with a relatively high doping level, significant impurity scattering rises, and the mobility can be much lower than the bulk [3]. In fact, there is no substantial evidence supporting that the carrier transport at the VLS grown silicon nanowire is quasi-ballistic.

### (5) Non-linearity accompanied by the rectification behavior

We noticed the apparent non-linearity in the rectifying IV curves (Figure. 3A of [1]) in the experimental results. The quasi-ballistic transport model does not provide a clear explanation for the non-linearity accompanied by the rectification behavior. While the non-linearity is typical and reasonable in the potential barrier dominated transport model.

## II. BIBL in the geometric diode.

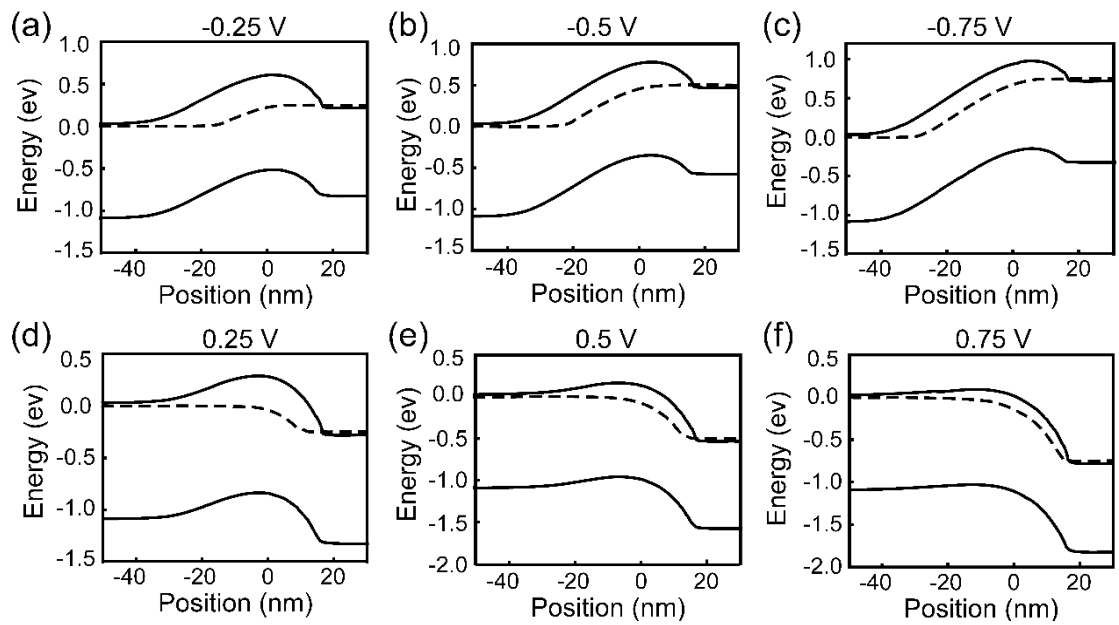

**FIG. S2. Band diagrams of geometric diodes.** Energy band diagrams with voltages of **(a)** -0.25 V, **(b)** -0.5 V, **(c)** -0.75 V, **(d)** 0.25 V, **(e)** 0.5 V, **(f)** 0.75 V, respectively.

## III. Depletion at the bottleneck of the geometric diode.

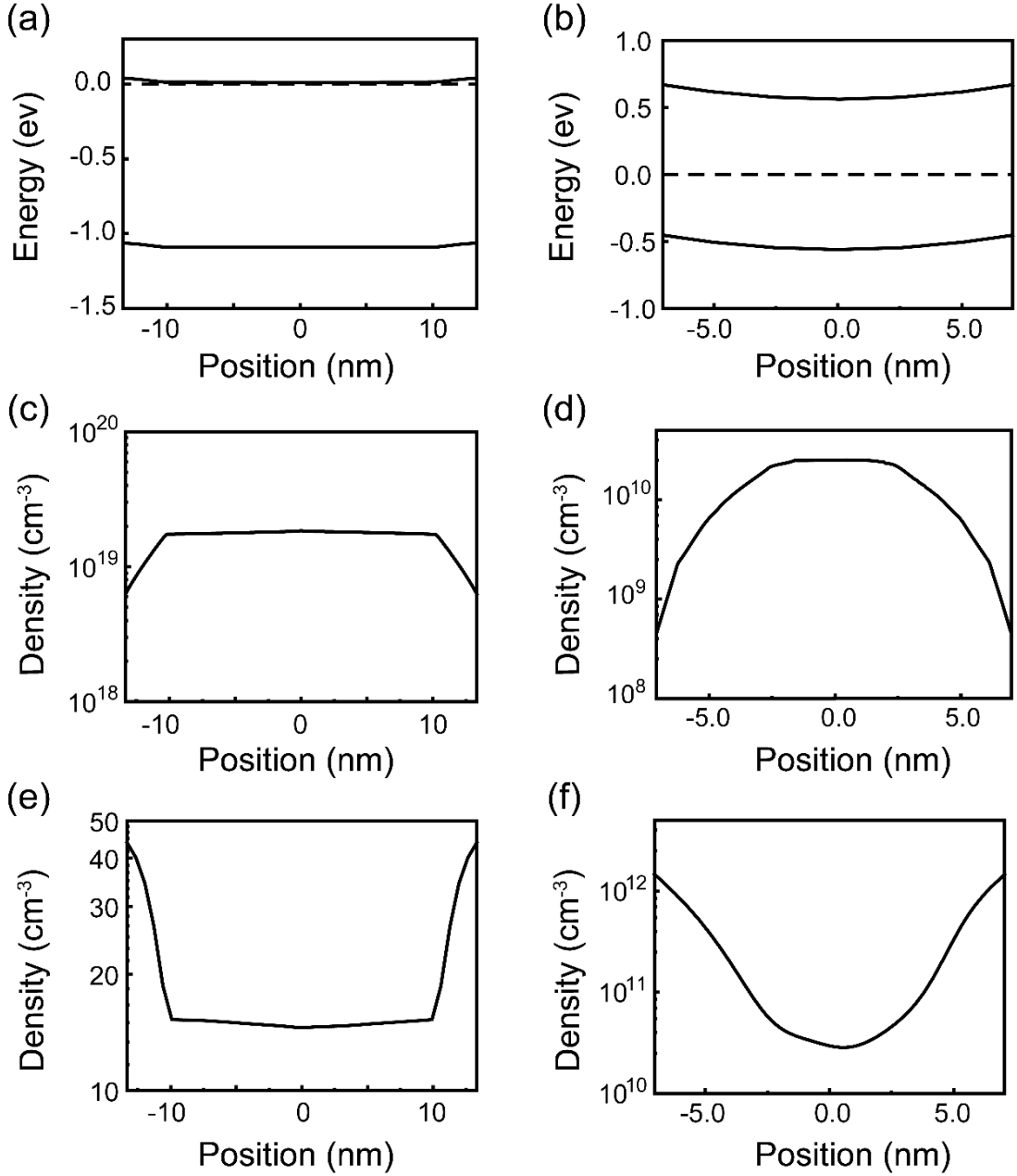

**FIG. S3. Band diagrams in the thick and bottleneck of geometric diode.** Energy band diagrams of geometric diode in the thick **(a)** and the bottleneck **(b)**, respectively. The electron density distribution in the thick **(c)** and the bottleneck **(d)** of geometric diode, respectively. The hole density distribution in the thick **(e)** and the bottleneck **(f)** of geometric diode, respectively.

## IV. The absence of minority accumulation in the geometric diode.

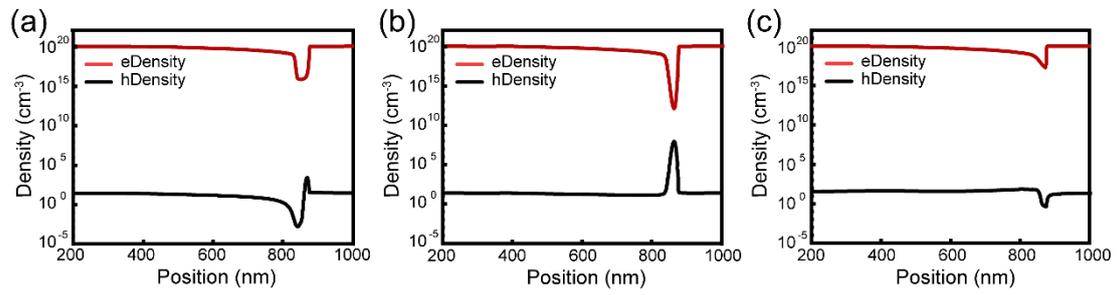

**FIG. S4. The electron and hole density distribution.** The electron and hole density distribution of silicon geometric nanowire diodes under the applied voltage of -1 V **(a)**, 0 V **(b)**, 1 V **(c)**, respectively.

## V. Surface state free geometric diode.

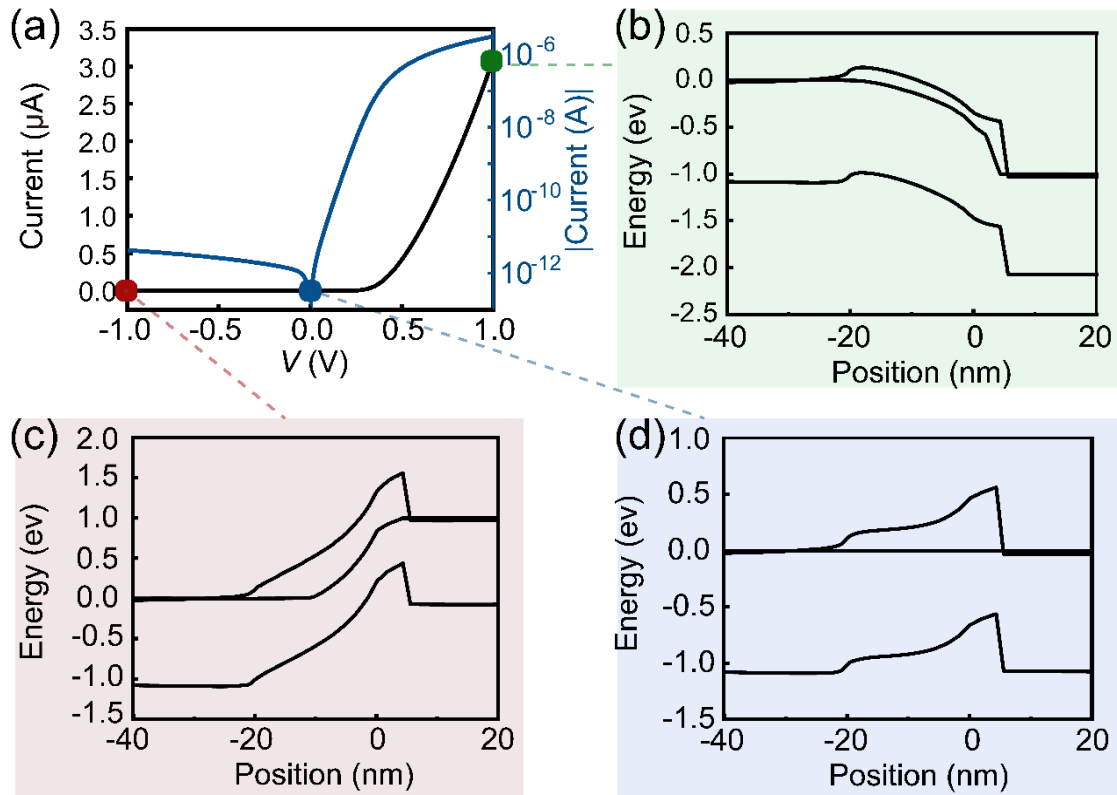

**FIG. S5. BIBL in surface state free geometric diode. (a)** The *I-V* curves of surface state free silicon geometric nanowire diodes with the undoped bottleneck. The energy band diagrams with voltages of **(b)** 1 V, **(c)** -1 V, **(d)** 0 V, respectively.

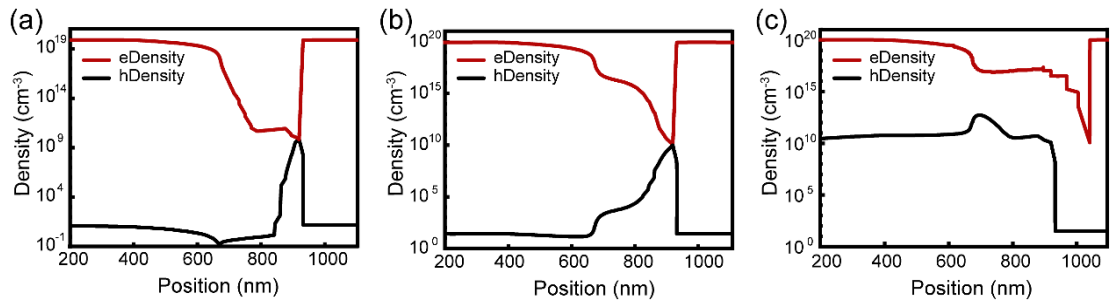

**FIG. S6. Carrier distribution of the surface state free geometric diode.** The electron and hole density distribution of silicon nanowire geometric diodes with the applied voltage of **(a)** -1 V, **(b)** 0 V, **(c)** 1 V, respectively.

## VI. Geometric dependence of the geometry diode.

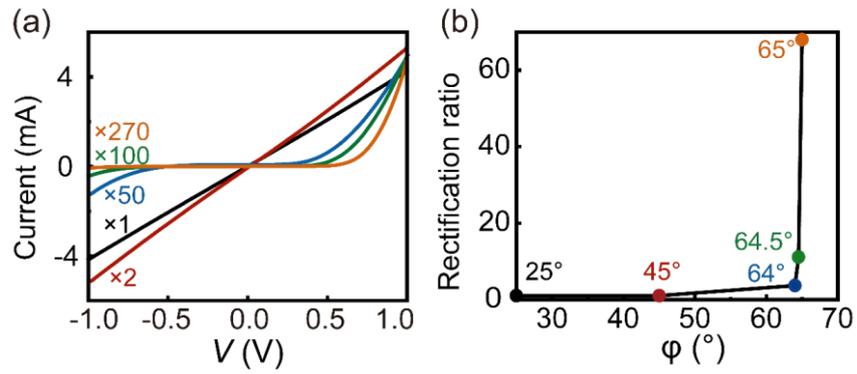

**FIG. S7. Geometric dependence of geometry diode**. **(a)** The *I-V* curves simulated from separate single-NW devices with $\varphi$ of 25°, 45°, 64°, 64.5°, 65°. **(b)** The extracted rectification ratio with various $\varphi$.

# VII. High-frequency rectification behavior.

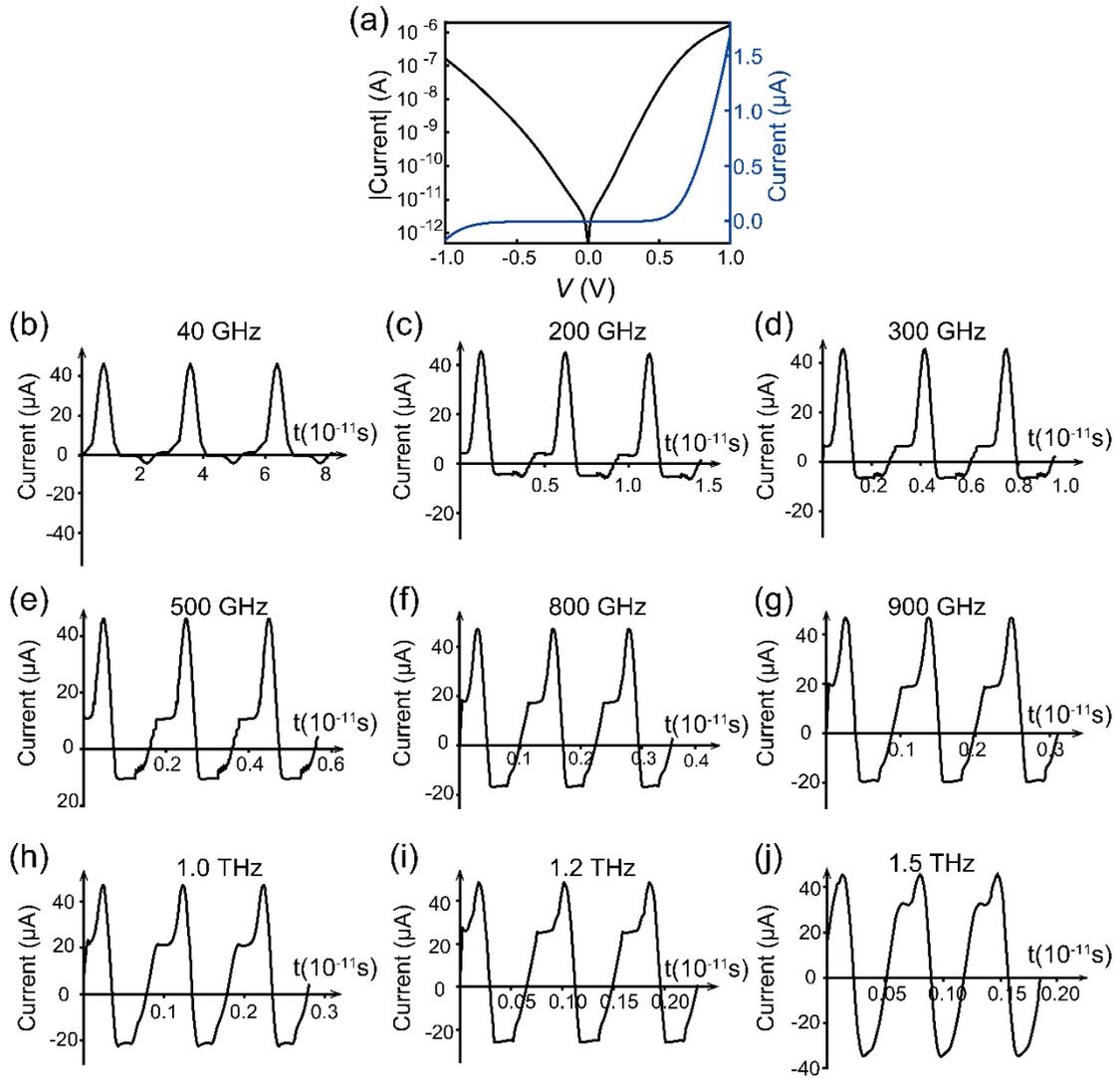

**FIG. S8. High-frequency response of silicon nanowire geometric diode. (a)** The static *I-V* of silicon nanowires geometric diode showing a rectification ratio of 10. Current response of the geometric diode with a rectification ratio of **(b)** 9.5 at 40 GHz **(c)** 8.5 at 200 GHz, **(d)** 6.9 at 300 GHz, **(e)** 4.2 at 500 GHz, **(f)** 2.7 at 800 GHz, **(g)** 2.4 at 900 GHz, **(h)** 2.2 at 1000 GHz, **(i)** 1.8 at 1.2 THz, and **(j)** 1.5 at 1500 GHz, respectively.

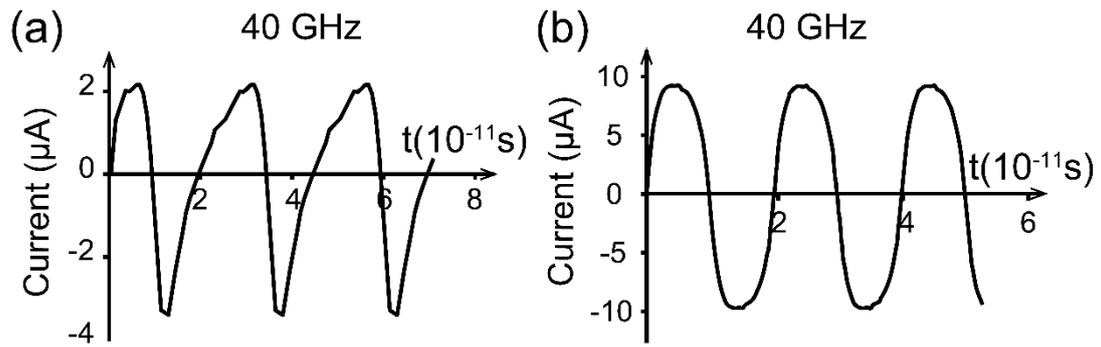

**FIG. S9. High-frequency response of p-n junction diode and Schottky junction diode.** The current v.s. time curve of a p-n junction diode **(a)** and Schottky junction diode **(b)** at 40 GHz.